\documentstyle[aps,twocolumn,epsf]{revtex}

\begin{document}
\title{       Efficient calculation of imaginary time 
               displaced correlation functions in the projector
               auxiliary field quantum Monte-Carlo algorithm}
\author{M. Feldbacher and F.F. Assaad }
\address{Institut f\"ur Theoretische Physik III, \\
   Universit\"at Stuttgart, Pfaffenwaldring 57, D-70550 Stuttgart, Germany.}
\maketitle

\begin{abstract}
The calculation of imaginary time displaced correlation functions with
the auxiliary field projector quantum Monte-Carlo algorithm
provides  valuable insight (such as spin and charge gaps) in  the model
under consideration.  One of the authors and M. Imada
\cite{Assaad96a} have proposed a 
numerically stable method to compute those quantities. Although precise   
this method is expensive in CPU time. Here, we present an alternative
approach which is an order of magnitude quicker, just as precise, and
very simple to implement. The method is based on the observation that for 
a given auxiliary field the equal time Green function matrix, 
$G$, is a projector: $G^2 = G$.  
\end{abstract}

    For a given Hamiltonian 
$ H = \sum_{x,y}  c_{x}^{\dagger}  T_{x,y} c_{y} +   H_I $ 
and its ground state $|\Psi_0 \rangle$, our aim is to calculate 
\begin{equation} 
 G^{<}_{x,y}(\tau) =
    \frac{ \langle \Psi_0 | c_{y}^{\dagger}(\tau)
c_{x} | \Psi_0 \rangle} {\langle \Psi_0 | \Psi_0 \rangle}, \; \; \tau \geq 0.
\end{equation}  
Here, $c_{x}^{\dagger}$ creates an electron with quantum numbers $x$,
$c_{x}(\tau) = e^{ \tau (H - \mu N) } c_{x} e^{-\tau(H - \mu N)}$,   
and the chemical potential $\mu  = E_0^{N+1} -  E_0^{N}$.
$ H_I$ corresponds to the interaction.
Within the  projector Quantum Monte Carlo (PQMC) algorithm, this quantity is
obtained by propagating a trial wave  $|\Psi_T \rangle$
function along the imaginary time axis \cite{Koonin86,Sorella89,Sorella89a}:
\begin{eqnarray}
\frac{\langle \Psi_0 | c_{y}^{\dagger}(\tau) c_{x} | \Psi_0 \rangle } 
{\langle \Psi_0 | \Psi_0 \rangle } &= & 
   \lim_{ \Theta \rightarrow \infty } \frac{
\langle \Psi_T |e^{-\Theta H } c_{y}^{\dagger}(\tau) c_{x} e^{-\Theta H } |
\Psi_T \rangle } { \langle \Psi_T |e^{-2\Theta H } | \Psi_T \rangle } \\
\nonumber
& \equiv & \lim_{ \Theta \rightarrow \infty } G^{<} (\Theta,\Theta + \tau ).
\end{eqnarray}
The above is valid provided that: $ \langle \Psi_0 |\Psi_T \rangle \neq 0 $.

To fix the notation, we will 
briefly summarize the essential steps required 
for the calculation of the right hand side (rhs) of the above equation at 
fixed values of the projection parameter $\Theta$. A detailed review may
be found in \cite{Loh92}. The formalism  - without numerical stabilization -
to compute time displaced correlation functions follows Ref. \cite{Deisz95}.
The first step is to carry out a Trotter
decomposition of the imaginary time propagation: 
\begin{equation}
       e^{-2\Theta H}=\left(
       e^{-\Delta \tau H_{t}/2}
       e^{-\Delta \tau H_{I}}
       e^{-\Delta \tau H_{t}/2}\right)^{m}+O((\Delta \tau )^{2}).
\label{Trotter}
\end{equation}
Here, $H_{t}$ ($H_{I}$) denotes the kinetic (interaction) term of the
model and $m\Delta \tau =2\Theta $. Having isolated the interaction
term, $H_{I}$, one may carry out a  Hubbard Stratonovitch (HS)
transformation to obtain: 
\begin{equation}
e^{-\Delta \tau H_{I}}=\sum_{\vec{s}} \exp \left( \sum_{x,y}c_{x}^{\dagger
}D_{x,y}(\vec{s})c_{y}\right) ,  \label{HS}
\end{equation}
where $\vec{s}$ denotes a vector of HS fields. 
For a Hubbard interaction, one can for example use various forms
of  Hirsch's discrete HS 
decomposition \cite{Hirsch83,Assaad98b}. For interactions taking 
the form of a perfect square,  decompositions presented in
\cite{Assaad97} are useful. 

The imaginary time propagation may now be written as: 
\begin{eqnarray}
&&e^{-2\Theta H}=\sum_{\vec{s}}U_{\vec{s}}(2\Theta ,0)+O((\Delta \tau )^{2})
 \; \;{\rm where}   \\
&&U_{\vec{s}}(2\Theta ,0)=\prod_{n=1}^{m}e^{-\Delta \tau
H_{t}/2} e^{\sum_{x,y}c_{x}^{\dagger }D_{x,y}(\vec{s}
_{n})c_{y}} e^{-\Delta \tau H_{t}/2}. \nonumber
\end{eqnarray}
The HS field has acquired an additional imaginary time 
index since we need independent fields for each time increment. 

The trial wave function is required to be a Slater determinant: 
\begin{equation}
|\Psi _{T}\rangle =\prod_{n=1}^{N_{p}}\left( \sum_{x}c_{x}^{\dagger
}P_{x,n}\right) |0\rangle .  \label{Trial}
\end{equation}
Here $N_{p}$ denotes the number of particles and $P$ is an $N_{s}\times N_{p}
$ rectangular matrix where $N_{s}$ is the number of single particle states.
Since $U_{\vec{s}}(2\Theta ,0)$ describes the propagation of non-interacting
electrons in an external HS field, one may integrate out the fermionic
degrees of freedom to obtain: 
\begin{equation}
G^{<} (\Theta, \Theta + \tau) =  
 \sum_{\vec{s}} W_{\vec{s}} G_{\vec{s}}^{<} (\Theta, \Theta) B_{\vec{s}} 
                                             (\Theta, \Theta + \tau) 
\label{G>}
\end{equation}
where we have omitted the $(\Delta \tau) ^2$ systematic error produced by
the Trotter decomposition. 
In the above equation, 
\begin{eqnarray*}
 B_{\vec{s}} && (\Theta _{2},\Theta _{1})= \\
&& \left\{
\begin{array}{c}
\prod_{n=n_{1}+1}^{n_{2}}e^{-\Delta
\tau T/2}e^{D(\vec{s_{n}})}e^{-\Delta \tau T/2}\;\; 
      {\rm if } \;\; \Theta _{2} > \Theta _{1} \\ 
 B^{-1}_{\vec{s}}(\Theta _{1},\Theta _{2}) \;\; 
{\rm if } \; \; \Theta _{1} >  \Theta _{2} 
\end{array}
\right.
\end{eqnarray*}
where $n_{1}\Delta \tau =\Theta _{1}$  and  $n_{2}\Delta \tau =\Theta
_{2},$
\begin{eqnarray*}
M_{\vec{s}}=P^{T}B_{\vec{s}}(2\Theta ,0)P,   
\; \; 
W_{\vec{s}}=\frac{\det (M_{\vec{s}})}{\sum_{\vec{s}}\det (M_{\vec{s}})}
\end{eqnarray*}
and 
\begin{eqnarray*}
G^{<}_{\vec{s}}(\Theta ,\Theta ) = 
  R_{\vec{s}}\left( \Theta \right) \left[ L_{\vec{s}}\left( \Theta \right) 
R_{\vec{s}}\left(
\Theta \right) \right] ^{-1}L_{\vec{s}}\left( \Theta \right), \\ 
R_{\vec{s}}\left( \Theta \right)  = B_{\vec{s}}(\Theta ,0)P, \; \; 
L_{\vec{s}}\left( \Theta \right)  =P^{T}B_{\vec{s}}(2\Theta ,\Theta )
\end{eqnarray*}

Restricting ourselves to models where $W_s$ is  positive definite (such
as the half-filled Hubbard, half-filled  Kondo lattice or attractive Hubbard
models) we can sample the probability distribution with Monte Carlo methods.  
For each auxiliary field configuration we then have to evaluate the 
quantity  $G_{\vec{s}}^{<} (\Theta, \Theta) B_{\vec{s}}
(\Theta, \Theta + \tau ) $ in a numerically stable and efficient way. 
This corresponds to the subject of the paper. 

At first glance it is clear that the evaluation of  
$G_{\vec{s}}^{<} (\Theta, \Theta) B_{\vec{s}}
(\Theta, \Theta + \tau ) $ is a numerically ill posed problem. We illustrate
this by considering free electrons on a
two-dimensional square lattice. 
\begin{equation}
H=-t\sum_{<\vec{i},\vec{j}>}c_{\vec{i}}^{\dagger }c_{\vec{j}}.
\end{equation}
Here, the sum runs over nearest-neighbors. For this Hamiltonian one has: 
\begin{equation}
    \langle \Psi _{0}|c_{\vec{k}}^{\dagger }(\tau )c_{\vec{k}}|\Psi _{0}\rangle
=  \langle \Psi _{0}|c_{
\vec{k}}^{\dagger }c_{\vec{k}}|\Psi _{0}\rangle  \exp \left( \tau (\epsilon _{\vec{k}}-\mu )\right),  \label{problem}
\end{equation}
where $\epsilon _{\vec{k}}=-2t(\cos (\vec{k}\vec{a}_{x})+\cos (\vec{k}\vec{a}%
_{y}))$, $\vec{a}_{x}$, $\vec{a}_{y}$ being the lattice constants.  We will assume $|\Psi_{0}\rangle $ to be non-degenerate. In a numerical calculation the
eigenvalues and eigenvectors of the above Hamiltonian will be known up to
machine precision, $\epsilon $. In the case $\epsilon _{\vec{k}}-\mu >0$, $%
\langle \Psi _{0}|c_{\vec{k}}^{\dagger }c_{\vec{k}}|\Psi _{0}\rangle \equiv 0
$. However, on a finite precision machine the later quantity will take a
value of the order of $\epsilon $. When calculating $\langle \Psi _{0}|c_{%
\vec{k}}^{\dagger }(\tau )c_{\vec{k}}|\Psi _{0}\rangle $ this roundoff error
will be blown up exponentially and the result for large values of $\tau $
will be unreliable.

In the PQMC approach and since for a given HS configuration, 
we have independent electrons in an external
field a similar form is obtained for the time displaced Green function. 
The $B_{\vec{s}}$ matrix plays the role of the exponential factors,
and contains 
exponentially large and small scales whereas 
$ G^{<}_{\vec{s}} \left( \Theta ,\Theta \right) $
contains scales bounded by order unity.
Since we equally  expect the result  
$ G^{<}_{\vec{s}} \left( \Theta ,\Theta +\tau \right) $ 
to be bounded 
by order unity, we will eventually run into numerical problems when
$\tau$ becomes {\it large}.

In order to circumvent this problem, Assaad and Imada \cite{Assaad96a}
proposed to 
do the calculation at finite
temperatures and then take the limit to vanishingly small temperatures. 
For the example of free electrons this amounts in doing calculation via: 
\begin{eqnarray}
    \langle \Psi_0 | c_{\vec{k}}^{\dagger}(\tau) c_{\vec{k}} | \Psi_0 \rangle =
     \lim_{\beta \rightarrow \infty} \frac
       { \exp \left(  \tau (\epsilon_{\vec{k}} - \mu)   \right) }
       { 1 +  \exp \left(  \beta (\epsilon_{\vec{k}} - \mu)   \right) }.
\end{eqnarray}
Even if the eigenvalues are known only up to machine precision, the right hand
side of the above equation for large but finite values of $\beta$ is
a numerically stable operation.    
To implement this idea in the QMC method,  Assaad and Imada 
considered a single particle 
Hamilton $H_0$ which has the trial wave function, $| \Psi_T \rangle $
as non-degenerate ground state and then compute:
\begin{eqnarray}
\label{ASM}
G_{\vec{s}}^{<} (\Theta,\Theta + \tau)&& \equiv \\
\lim_{\beta \rightarrow \infty} & & 
\frac{
      {\rm Tr} \left( e^{-\beta H_0}  U_{\vec{s}}(2 \Theta,\Theta)
                     c_{y}^{\dagger}(\tau) c_{x}
                     U_{s}(\Theta,0) \right) } 
     { {\rm Tr} \left( e^{-\beta H_0} U_{s}(2 \Theta,0)  \right) }
\nonumber.
\end{eqnarray}
Although the rhs of the above equation may be computed in a numerically stable 
way, the approach is cumbersome and numerically expensive. In particular, 
for each measurement, all quantities have to be computed from scratch since
the ad-hoc inverse temperature $\beta$ has to be taken into account. 

Here, we propose an alternative method. 
We will again start with the example of
free electrons. Since, 
$\langle \Psi_0 | c_{\vec{k}}^{\dagger}(\tau) c_{\vec{k}} |  \Psi_0 \rangle 
=1,0 $, we can rewrite Eq. \ref{problem} as:
 
\begin{equation}
    \langle \Psi _{0}|c_{\vec{k}}^{\dagger }(\tau )c_{\vec{k}}|\Psi _{0}\rangle
=  \left( \langle \Psi _{0}|c_{
\vec{k}}^{\dagger }c_{\vec{k}}|\Psi _{0}\rangle 
 \exp \left(  (\epsilon _{\vec{k}}-\mu )\right) \right)^{\tau} 
\label{problemsolved}
\end{equation}
which involves only 
well defined numerical manipulations even in the large $\tau$ limit.

The implementation of this idea in the QMC algorithm is as follows. First, one
has to notice that the Green function $G^{<}_{\vec{s}}
(\Theta, \Theta)$ is a projector:
\begin{equation}
\label{Proj}
G^{<}_{\vec{s}}(\Theta, \Theta)^2 = G^{<}_{\vec{s}}(\Theta, \Theta).
\end{equation}
  This is simply shown
by carrying out a singular value decomposition of the 
$N_{s}\times N_{p}$  $R_{\vec{s}}\left( \Theta \right) $ and 
$L_{\vec{s}}\left(
\Theta \right) $ matrices  
\begin{eqnarray*}
R_{\vec{s}}\left( \Theta \right)  &=&U_{r,\vec{s}} D_{r,\vec{s}}
   V_{r,\vec{s}} \\
L_{\vec{s}}\left( \Theta \right)  &=&V_{l,\vec{s}} D_{l,\vec{s}}
    U_{l,\vec{s}}.
\end{eqnarray*}
Here $U_r$ ($U_l$) 
is a $N_{s}\times N_{p}$ ($N_{p}\times N_{s}$) and column (row) 
orthogonal matrix, $D_{l,r}$  are diagonal 
$N_{p}\times N_{p}$ matrices and  $V_{l,r}$ unit upper triangular $N_{p}\times N_{p}$ matrices.
For the equal time Green function  only $U_{r,\vec{s}},U_{l,\vec{s}}$
are important 
\begin{eqnarray*}
G^{<}_{\vec{s}}(\Theta ,\Theta )=U_{r,\vec{s}}\left[ U_{l,\vec{s}}U_{r,\vec{s}}\right] ^{-1}U_{l,\vec{s}}.
\end{eqnarray*}
Eq.  \ref{Proj} then follows from:
\begin{eqnarray*}
\left(G^{<}_{\vec{s}} (\Theta ,\Theta ) \right)^2 & &
= U_{r,\vec{s}}\left[ U_{l,\vec{s}}U_{r,\vec{s}}\right]
^{-1}\left[ U_{l,\vec{s}}U_{r,\vec{s}}\right] \left[ U_{l,\vec{s}}U_{r,\vec{s}}\right] ^{-1}U_{l,\vec{s}} \\
& & =
G^{<}_{\vec{s}}(\Theta ,\Theta ).
\end{eqnarray*}
This in turn implies that $G^{<}_{\vec{s}}(\Theta_1,\Theta_3)$ obeys a 
simple composition identity
\begin{eqnarray}
\label{composition}
G^{<}_{\vec{s}}(\Theta_1,\Theta_2)G^{<}_{\vec{s}}(\Theta_2,\Theta_3) =
G^{<}_{\vec{s}}(\Theta_1,\Theta_3)
\end{eqnarray}
since
\begin{eqnarray*}
G^{<}_{\vec{s}}(\Theta_1,\Theta_3)  = 
G^{<}_{\vec{s}}\left( \Theta_1,\Theta_1 \right)  & &
      B_{\vec{s}}\left( \Theta_1,\Theta_3\right) = 
\\
\left(G_{\vec{s}}^{<}\left( \Theta_1,\Theta_1\right) \right)^2 
B_{\vec{s}}\left( \Theta_1,\Theta_3\right)  &=& 
G^{<}_{\vec{s}}\left( \Theta_1,\Theta_1\right) 
G_{\vec{s}}^{<}(\Theta_1,\Theta_3) 
\\
=G^{<}_{\vec{s}}(\Theta_1,\Theta_2)G^{<}_{\vec{s}}(\Theta_2,\Theta_3). & &
\end{eqnarray*}

Using this  composition property (\ref{composition}) we can break up a large $%
\tau $ interval into a set of smaller intervals of length $\tau = N \tau_1$
so that
\begin{equation}
G^{<}_{\vec{s}}\left( \Theta ,\Theta +\tau \right) = 
\prod_{n =0}^{N-1} G^{<}_{\vec{s}} \left( \Theta + \left[ n+1 \right] \tau_1
,\Theta +n\tau_{1}\right)   \label{mult K}
\end{equation}
The above equation is the generalization of Eq. \ref{problemsolved}.
If $\tau_1$ is  {\it small} enough each Green function in
the above product is accurate and has matrix elements bounded by order unity. 
The matrix multiplication is then numerically well defined.   

We illustrate the efficiency of the method for the Kondo lattice model:

\begin{equation}
H_{KLM} =
 -t \sum_{\langle \vec{i},\vec{j} \rangle ,\sigma }
 c^{\dagger}_{\vec{i},\sigma} c_{\vec{j},\sigma}
    + J \sum_{\vec{i}}
    \vec{S}^{c}_{\vec{i}} \vec{S}^{f}_{\vec{i}}.
\end{equation}
Here $\vec{i}$ runs over the $L^2$-sites of a square lattice,
$\langle \vec{i},\vec{j} \rangle$ corresponds to nearest neighbors,
$c^{\dagger}_{\vec{i},\sigma} $
creates a conduction electron with
z-component of spin $\sigma$  on site $\vec{i}$ and  periodic
boundary conditions are imposed.
$ \vec{S}^{f}_{\vec{i}} =(1/2) \sum_{\sigma,\sigma'}
f^{\dagger}_{\vec{i},\sigma}    \vec{\sigma}_{\sigma,\sigma'}
f_{\vec{i},\sigma'} $ with
$\vec{\sigma}$ the Pauli matrices.
An equivalent form holds  for the conduction electrons.
A constraint of one fermion per $f$-site is enforced.
As shown in Ref. \cite{Assaad99a} at half-filling, the PQMC method
may  be used to carry out sign-free simulations of the model.
\begin{figure}[h]
\epsfxsize=8.8cm
\hfil\epsfbox{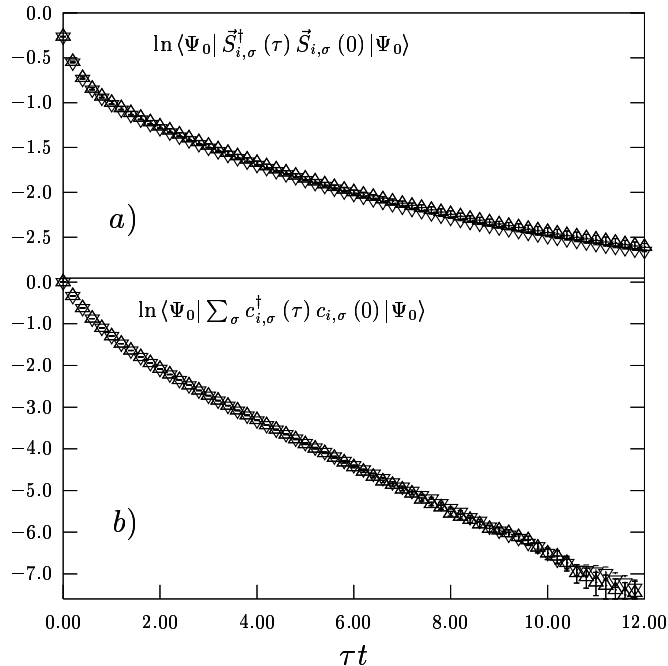}\hfil
\caption[]
{\noindent Imaginary time displaced  on-site spin-spin (a) and Green function
(b) correlation function. We consider a $6 \times 6 $ lattice at half-filling
and $J/t =1.2$. In both (a) and (b) results obtained form Eq. (\ref{mult K})
($ \bigtriangleup $) and (\ref{ASM}) ($ \bigtriangledown $) are plotted.
\label{Comp.fig} }
\end{figure}
Fig. \ref{Comp.fig} plots the on-site time displaced spin-spin 
correlation functions as well as the on-site Green function
for a $ 6 \times 6 $ lattice at $J/t = 1.2$ and
half-band filling.  Here, we consider the total spin: $\vec{S}_{\vec{i}}
=  \vec{S}^{f}_{\vec{i}} + \vec{S}^{c}_{\vec{i}}$.  Both methods based on Eq.
(\ref{mult K}) and Eq. (\ref{ASM})
produce identical results within the error-bars.
(Had we used the same series of random
numbers, we would have obtained exactly the same values up to roundoff errors
which are of the order  $10^{-8}$)

The important point however, is that the method based on Eq. (\ref{mult K}),
for this special case, more than an order of magnitude quicker  in CPU
time than the calculation based on Eq. (\ref{ASM}). 
A calculation following Eq. (\ref{mult K}) involves matrix inversions 
(multiplications) of size $N_p \times N_p$ ( ($N_p \times N) 
\cdot (N \times N) $). Here,  $N$ denotes the number of sites. To
this we have add that many quantities required for the calculation 
are at hand during 
the simulation and do not have to be recalculated.
On the other hand, the method based on Eq. (\ref{ASM}) involves 
matrix inversions and multiplications of size  up to  $ 2 N \times 2 N$
\cite{Assaad96a}.
In this approach and 
for given set of HS fields all quantities have to be computed from scratch.

In summary, we have  described an efficient method for the calculation
of imaginary time displaced correlation functions in the framework 
of the  PQMC algorithm.  The method is elegant and easy to implement in
a standard PQMC code and is an order of magnitude quicker than previously
used methods. We have demonstrated  the efficiency of the method in the 
special case
of the two dimensional Kondo lattice model. Given the ability of calculating
efficiently 
time displaced correlation functions  at arbitrarily large  imaginary
times enables  us to pin down charge and spin gaps \cite{Assaad99a}
 as well as 
quasiparticle weights  \cite{Brunner00b}.  Dynamical propertied may equally
be obtained after continuation to real time via the Maximum Entropy method
\cite{Jarrell96}.

We acknowledge S. Capponi for useful conversations. The calculation were
carried out on the Cray T3E of the HLRS (Stuttgart).  M. Feldbacher 
the DFG  for financial support, grant number MU 820/10-1.

\end{document}